\documentclass[twocolumn,superscriptaddress]{revtex4-2}
\usepackage{amsmath}
\usepackage{amssymb}
\usepackage{bm}
\usepackage{epsfig}
\usepackage{graphicx}
\usepackage{color}
\usepackage[unicode=true,colorlinks=true,citecolor=blue,urlcolor=blue]{hyperref} 
\usepackage[utf8]{inputenc}
\usepackage[english]{babel}
\usepackage{xcolor}
\usepackage{gensymb}
\usepackage{physics}
\usepackage{array}[=2016-10-06]

\usepackage{float}

\usepackage{multirow}

\newcommand{\tsub}[1]{\texorpdfstring{\textsubscript{#1}}{#1}}

\begin{document}
\title{Structural phase transitions in double perovskite crystals\\ studied by Brillouin light scattering}

\author{D.~O.~Horiachyi}
\email[correspondence address: ]{dmytro.horiachyi@tu-dortmund.de}
\affiliation{Experimentelle Physik 2, Technische Universit\"at Dortmund, 44227 Dortmund, Germany}
\author{M.~O.~Nestoklon}
\affiliation{Experimentelle Physik 2, Technische Universit\"at Dortmund, 44227 Dortmund, Germany}
\author{I.~A.~Akimov}
\affiliation{Experimentelle Physik 2, Technische Universit\"at Dortmund, 44227 Dortmund, Germany}
\author{D.~R.~Yakovlev}
\affiliation{Experimentelle Physik 2, Technische Universit\"at Dortmund, 44227 Dortmund, Germany}
\author{V. Vasylkovskyi}
\affiliation{Experimental Physics VI, Julius-Maximilian University of W\"urzburg, 
97074 W\"{u}rzburg, Germany}
\author{O.~Trukhina}
\affiliation{Experimental Physics VI, Julius-Maximilian University of W\"urzburg, 
97074 W\"{u}rzburg, Germany}
\author{V.~Dyakonov}
\affiliation{Experimental Physics VI, Julius-Maximilian University of W\"urzburg, 
97074 W\"{u}rzburg, Germany}
\author{M.~Bayer}
\affiliation{Experimentelle Physik 2, Technische Universit\"at Dortmund, 44227 Dortmund, Germany}
\affiliation{Research Center FEMS, Technische Universit\"{a}t Dortmund, 44227 Dortmund, Germany}

\begin{abstract}

Inorganic lead-free double perovskites represent particular interest as non-toxic and stable material platform for optoelectronic applications. Here, we employ Brillouin light scattering spectroscopy to investigate the elastic properties and structural phase transitions in single crystals of Cs$_2$AgBiBr$_6$ and Cs$_2$AgBiCl$_6$. A complete set of elastic constants is determined from the Brillouin scattering measurements performed along three different crystallographic directions. Both materials exhibit similar elastic constants and weak elastic anisotropy in the cubic phase. At low temperatures, the lifting of degeneracy of transverse acoustic phonon modes is attributed to a lowering of crystal symmetry. From the temperature dependence of the acoustic phonon frequencies, we determine the structural phase transition temperature of about 43~K for Cs$_2$AgBiCl$_6$, compared to 122~K for the cubic-to-tetragonal phase transition in Cs$_2$AgBiBr$_6$.

\end{abstract}

%\date{\today}
\maketitle

\section{Introduction}
\label{sec:intro}

In recent years great interest to metal halide perovskites has grown rapidly due to the success in photovoltaic applications. The impressive results with power conversion efficiency of lead halide perovskites exceeded 26\% thanks to improving the device fabrication routines and the proper choice of atomic elements and their composition in perovskite material~\cite{Jia2025,Dyakonov2021}. The most intensively studied materials are lead halide perovskites with the general formula APbX$_3$, where A is an organic (MA: CH$_3$NH$_3$, FA: H$_2$NCHNH$_2$) or inorganic cation (Cs) and X is a halide anion (I, Br, or Cl). At the same hand the spectrum of perovskite semiconductors is very wide and it is not limited by lead-based perovskites. Lead can be replaced by other divalent metals such as Sn and Ge. Moreover, a new class of lead-free double perovskite semiconductors with the general formula A$_2$B$^{\prime}$B$^{\prime\prime}$X$_6$ with B$^{\prime}$ and B$^{\prime\prime}$ being monovalent and trivalent metal ions attracted significant attention due to higher stability and lower toxicity ~\cite{Tress2022,Zhang2022}.

Halide perovskites, as semiconductor materials, are of particular interest for fundamental research due to the strong interaction of charge carriers with the lattice, which gives rise to pronounced peculiarities in their electronic, optical, and elastic properties~\cite{Yamada2022}. Photoacoustic studies are of special interest in this context: strain can modify the band gap and exciton energies, while optical excitation with femtosecond pulses can induce coherent lattice vibrations~\cite{Maris1986,Young2012,Ruello2015}. Perovskites provide an appealing platform for such studies due to their rich phase diagram, soft acoustic phonons, and tunable excitonic properties~\cite{Nelson1994,Schade2019,Steele2020,Kirstein2020}. Moreover, efficient generation of high-frequency sub-THz coherent acoustic phonons enables ultrafast control of optical and electronic properties, as studied using femtosecond pump–probe spectroscopy, opening new opportunities for halide perovskites in hypersonics~\cite{Lejman2014,Mante2017,Shaller2017}. In particular, recent studies demonstrated that efficient generation of transverse acoustic phonons in the double Cs$_{2}$AgBiBr$_{6}$ perovskite occurs due to anisotropic photostriction in the low-temperature tetragonal phase~\cite{Horiachyi2025}. In this respect, knowledge of the elastic constants of various halide perovskites and their temperature dependence is crucial for photoacoustic studies.

Brillouin light scattering (BLS) spectroscopy is a powerful tool that has been applied, together with inelastic neutron scattering, for the evaluation of elastic softness in a wide range of lead halide perovskites~\cite{Ferreira-2018}. BLS has also been applied to the double perovskite Cs$_{2}$AgBiBr$_{6}$, which is among the most extensively studied double perovskites. The elastic constants of this material have been determined using both nanoindentation and BLS spectroscopies~\cite{Lun2022,BLS-2025}. However, most studies have been performed at room temperature and for the cubic phase, despite the fact that BLS is highly effective for monitoring structural phase transitions through the temperature dependence of the acoustic phonon peak positions~\cite{Horiachyi2025}. Other double perovskites, such as Cs$_{2}$AgBiCl$_{6}$, have been far less studied~\cite{BLS-2022,BLS-2025}.

In this work, we study the double perovskite semiconductors Cs$_{2}$AgBiBr$_{6}$ and Cs$_{2}$AgBiCl$_{6}$ using Brillouin light scattering spectroscopy. We evaluate and compare the elastic constants of both materials by measuring BLS from different crystallographic facets and at multiple wavelengths. Our results reveal weak anisotropy and a high similarity in the elastic properties of chloride and bromide compounds. At a low temperature of 5~K, we observe a lowering of cubic symmetry which is manifested in lifting the degeneracy of transverse acoustic phonons. From temperature dependence we determine the structural phase transition temperature to be 43~K for Cs$_{2}$AgBiCl$_{6}$, compared to 122~K for cubic to tetragonal phase transition in Cs$_{2}$AgBiBr$_{6}$.

\section*{Methods}
{\bf Crystal growth.}
The Cs$_{2}$AgBiBr$_{6}$ and Cs$_{2}$AgBiCl$_{6}$ crystals were grown using a controlled cooling crystallization technique in acidic media. High-purity precursors were used without additional purification: cesium bromide (CsBr, 99.999\%), cesium chloride (CsCl, 99.9\%), silver chloride (AgCl, 99,999\%), and bismuth trichloride (BiCl$_{3}$, 98\%) from “Sigma-Aldrich”; Silver bromide (AgBr, 99\%) and bismuth tribromide (BiBrl$_{3}$, 98\%) from “Alfa”. As solvents and crystal growth media,  hydrobromic acid (HBr, 48\%) from “Acros Organics” and hydrochloric acid (HCl, 36\%) from “ThermoScientific” were used for Cs$_{2}$AgBiBr$_{6}$ and Cs$_{2}$AgBiCl$_{6}$, respectively. 
Stoichiometric amounts of the precursors were dissolved in the respective acid by heating the mixture to 120~$^{\circ}$C for Cs$_{2}$AgBiBr$_{6}$ and 160~$^{\circ}$C for Cs$_{2}$AgBiCl$_{6}$ and maintained at the set temperature for 5 hours to ensure complete dissolution. The solution was gradually cooled at a rate of 1~$^{\circ}$C per hour to ambient temperature, promoting slow nucleation and crystal growth. The harvested crystals were washed with dichloromethane and dried in air.
 
{\bf Crystal characterization.} 
Phase purity and crystal structure were assessed by powder X-ray diffraction (XRD) using a General Electric XRD 3003 TT system equipped with a monochromatic copper (Cu)K$\alpha$ radiation source. The single crystals were manually pulverized and the resulting powders were uniformly spread onto a low-background polymer holder. Measurements were performed under ambient conditions in Bragg-Brentano geometry within the 10 - 70 2$\Theta^{\circ}$ range with a step size of 0.005$^{\circ}$ and an integration time of 5 seconds per step. The lattice parameters were refined from the indexed reflections assuming the cubic double perovskite structure (space group $Fm\bar{3}m$).

{\bf Brillouin light scattering.} 
Brillouin light scattering (BLS) was measured using a stabilized multipass double Fabry-Per\'ot Brillouin spectrometer (TFP-2 from Table Stable). The distance between the mirrors was set to 3 mm and the width of the entrance and exit slits was 450~$\mu$m. The excitation of the samples was performed with continuous wave ({\it cw}) single frequency lasers with photon energy of 2.287~eV (542~nm) and 2.407~eV (515~nm) with the spectral width below 40~neV (10~MHz). Microscope objective with 20$\times$ magnification and numerical aperture of 0.4 was used for focusing of the exciation laser with intensity of 3~mW into a spot with diameter of about 5~$\mu$m. The scattered light was collected by the same microscope objective in back-scattering geometry. The detection had the same linear polarization as the excitation laser, detecting only vertically polarized light. The position of the peaks is independent on the linear polarization direction with respect to crystallographic axes, which excludes possible birefringence. Below in the analysis of Brillouin spectra we assume that the refractive index is a constant independent on the direction. The samples were placed in a helium flow cryostat allowing the temperature dependent measurements from room temperature (RT) $T=293$~K down to 5~K. 
As-grown bulk single crystals of Cs$_{2}$AgBiCl$_{6}$ and Cs$_{2}$AgBiBr$_{6}$ contain only \{111\} facets. For Brillouin light scattering measurements along different crystallographic orientations, the crystals were subsequently polished to produce surfaces with normals along $\left\langle 001 \right\rangle$ and $\left\langle 110 \right\rangle$ directions.

\section{Experimantal results}
\label{sec:results}

\subsection{Cubic phase at room temperature}
\label{sec:RT}

\begin{figure}
\centering
  \includegraphics[width=.9\linewidth]{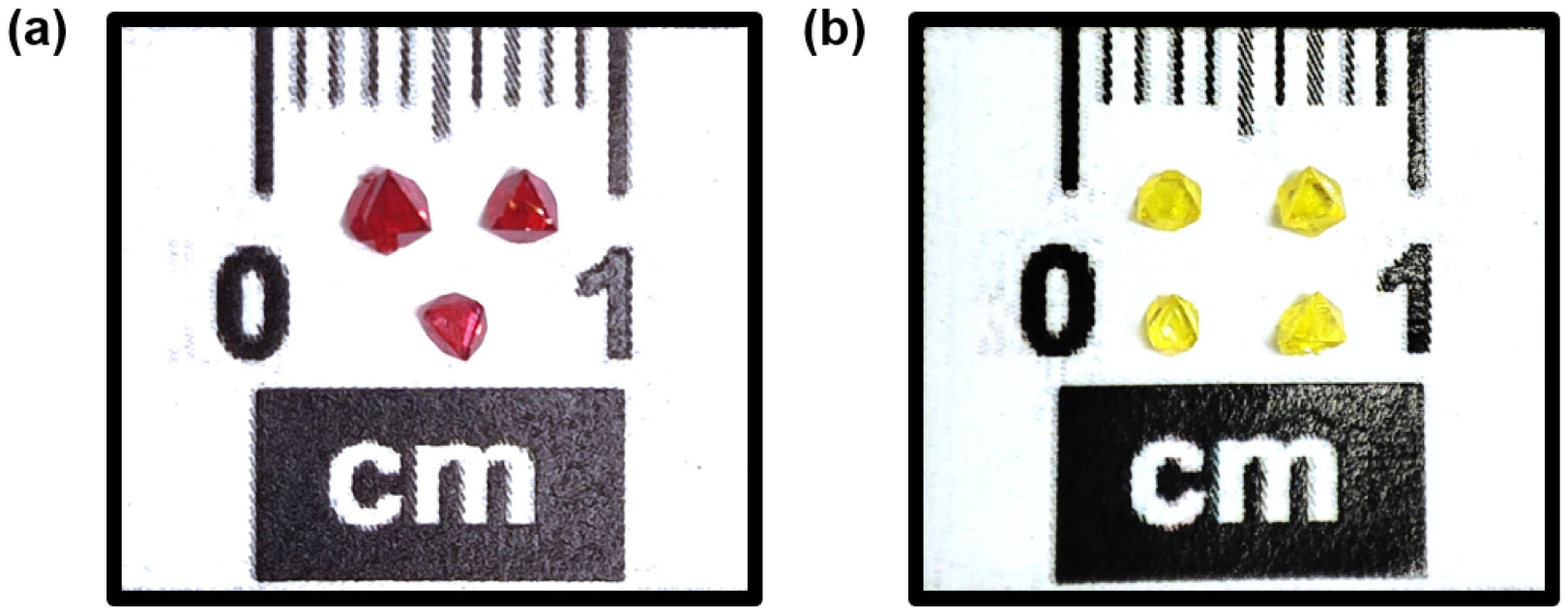}
  \caption{Optical images of (a) Cs$_{2}$AgBiBr$_{6}$ and (b) Cs$_{2}$AgBiCl$_{6}$ single crystals. The as-grown crystals are bounded solely by \{111\} facets.}
  \label{fig:samples}
\end{figure}

The controlled cooling growth yielded millimeter-sized single crystals with well-defined polyhedral shapes with prominent \{111\} facets for both compounds. Cs$_{2}$AgBiBr$_{6}$ crystals possess orange-red color (Figure~\ref{fig:samples}a), while Cs$_{2}$AgBiCl$_{6}$ crystals appear yellow (Figure~\ref{fig:samples}b), which is related to the increase of the energy bandgap during transition from Br to Cl halide anions ~\cite{Gray2019, Zelewski-2019, Jain-2025}. The average size of the chloride crystals was approximately half that of the bromide crystals.

\begin{figure}
\centering
  \includegraphics[width=.99\linewidth]{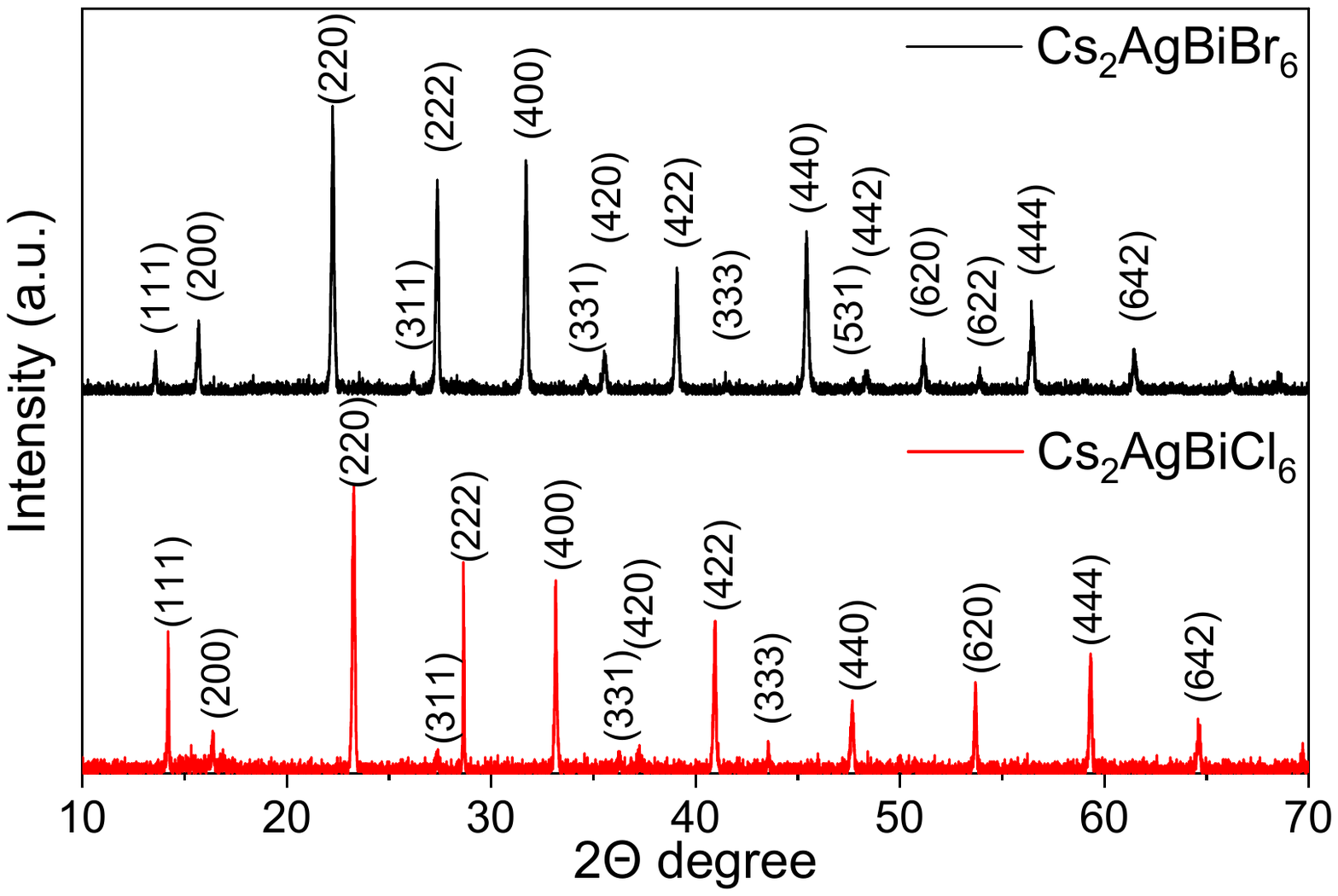}
  \caption{Powder X-ray diffraction patterns of Cs$_{2}$AgBiBr$_{6}$ and Cs$_{2}$AgBiCl$_{6}$ measured at room temperature $T=293$~K.}
  \label{fig:xrd}
\end{figure}

Powder XRD patterns confirmed the formation of a pure cubic double perovskite phase for both materials, with no impurity peaks detected as shown in Figure~\ref{fig:xrd}. Refinement of multiple indexed peaks yielded average lattice constants of $a = 11.29$~\AA{} for Cs$_{2}$AgBiBr$_{6}$ and $a = 10.79$~\AA{} for Cs$_{2}$AgBiCl$_{6}$, which aligns well with data reported in the literature~ \cite{McClure-2016, Gray2019}. The larger lattice in the bromide analog may reflect the greater ionic radius of Br- (1.96 \AA{}) compared to Cl- (1.81 \AA{}), which leads to the expanded [AgX$_{6}$] and [BiX$_{6}$] octahedra. 

\begin{figure}
\centering
\includegraphics[width=.99\linewidth]{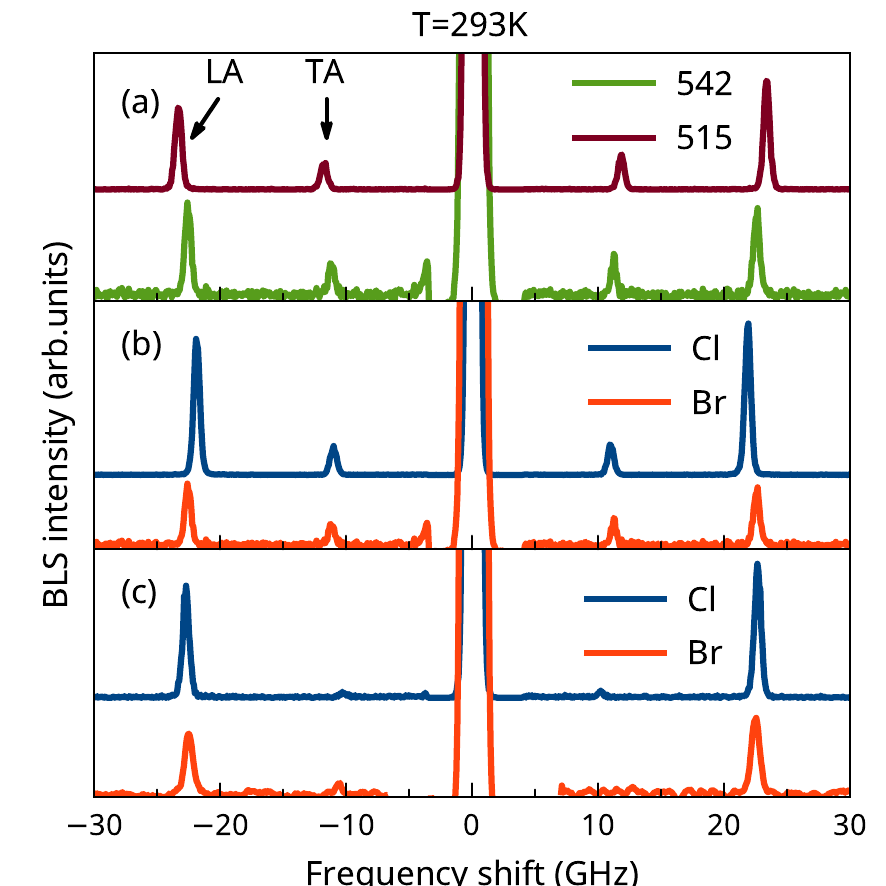}
  \caption{Brillouin light scattering (BLS) spectra measured at room temperature $T=293$~K. (a) Excitation with 542~nm (green) and 515~nm (brown) of Cs$_{2}$AgBiCl$_{6}$ along $\langle111\rangle$ crystalographic axis. BLS spectra under excitation with 542~nm laser in Cs$_{2}$AgBiCl$_{6}$ (blue) and Cs$_{2}$AgBiBr$_{6}$ (red) measured at room temperature. The data are shown in (b) and (c) for excitation along $\langle111\rangle$ and $\langle110\rangle$ crystallographic directions, respectively.}
  \label{fig:rtbls}
\end{figure}

The room temperature BLS data in Cs$_{2}$AgBiCl$_{6}$ and Cs$_{2}$AgBiBr$_{6}$ are summarized in Figure~\ref{fig:rtbls}. The BLS spectra comprise one or two narrow peaks with a spectral width of 0.7~GHz, which is limited by the spectral resolution of the setup. The lower energy peak centered at about 11~GHz is attributed to light scattering with emission or absorption of transverse acoustic (TA) phonon (Stokes or anti-Stokes peaks in BLS spectra, respectively), while the higher energy peak around 22~GHz corresponds to scattering on longitudinal acoustic (LA) phonon. Both Cs$_{2}$AgBiCl$_{6}$ and Cs$_{2}$AgBiBr$_{6}$ crystals exhibit the cubic structural phase at room temperature. In this case TA phonons are doubly degenerate and represented by one peak only. The spectral positions of the peaks depend on the composition of the double perovskite crystal (chloride vs bromide), excitation wavelength, and orientation of the crystals. 

BLS spectra in Cs$_{2}$AgBiCl$_{6}$ for excitation with different laser lines of 542 and 515~nm are shown in Fig.~\ref{fig:rtbls}(a). The data correspond to excitation along $\langle111\rangle$ crystallographic direction. In both cases we observe scatering on both TA and LA phonons. For the 542~nm laser we observe two peaks with frequencies of $21.88$~GHz and $11.01$~GHz, while for the 515~nm laser the BLS peaks shift to higher frequencies, i.e. for longitudinal acoustic phonon to $23.34$~GHz and for transverse acoustic phonons to $11.80$~GHz, respectively. This change is expected due to momentum conservation. The position of the peak is defined by the acoustic phonon velocity which is constant near the center of the Brillouin zone. In the back-scattering geometry, the velocity of the $s$-th mode may be calculated from the peak in BLS spectrum frequency $\delta \nu_s$ as:
\begin{equation}\label{eq:Vs}
  V_s = \frac{\lambda}{2n_r}{\delta \nu_s}\,,
\end{equation}
where $\lambda$ is the excitation wavelength and $n_r$ is the refractive index of material. For Cs\tsub{2}AgBiBr\tsub{6} refractive index for $\lambda=542$~nm is 2.17 and for $\lambda=515$~nm is 2.21 \cite{Polyanskiy2024}. For Cs\tsub{2}AgBiCl\tsub{6} we use 1.95 and 1.91 for two wavelengths used in experiment. We set the refractive index at larger wavelength to the value from Fig.3(b) in Ref.~\cite{Jain-2025} and their ratio to satisfy Eq.~\eqref{eq:Vs}.

  Comparison of BLS spectra in Cs$_{2}$AgBiCl$_{6}$ and Cs$_{2}$AgBiBr$_{6}$ under excitation with the same laser (542~nm) along  $\langle111\rangle$ (b), and $\langle110\rangle$ (c) crystallographic directions is presented in Fig.~\ref{fig:rtbls}(b) and (c), respectively. Similar to Cs$_{2}$AgBiCl$_{6}$, we observe two peaks with larger frequency shifts of $22.57$~GHz and $11.20$~GHz in Cs$_{2}$AgBiBr$_{6}$ along $\langle111\rangle$ crystallographic direction. For excitation  along $\langle110\rangle$ crystallographic direction we observe only one LA peak for chloride and bromide crystals with frequencies of $22.71$~GHz and $22.50$~GHz, respectively.  The data with phonon scattering frequencies $\delta\nu_s$, measured for different crystallographic directions and corresponding to BLS performed at 542~nm are summarized in Table~\ref{table:PhFreq}. It follows that the relative position of peak maxima for Cs$_{2}$AgBiCl$_{6}$ and Cs$_{2}$AgBiBr$_{6}$ measured along different crystallographic directions is close to each other. As shown below, since both the density and the value of refractive index are lower in Cs$_{2}$AgBiCl$_{6}$, the similar positions of BLS peaks indicate similar values of the elastic constants for both materials.

\begin{table*}[t]
\begin{tabular}{|c|p{3cm}|p{4cm}| p{4cm}|} 
 \hline
 Crystal facet & Acoustic phonon & Phonon frequency (GHz) \newline for Cs$_{2}$AgBiCl$_{6}$ & Phonon frequency (GHz) \newline for Cs$_{2}$AgBiBr$_{6}$  \\  
 \hline
 \multicolumn{4}{|c|}{Cubic phase}  \\ 
 \hline
  $\mathcal{S}_2$ \{100\} & LA &  22.70 & 22.82\\ 
 \hline
  $\mathcal{S}_1$ \{110\} & LA &  22.71 & 22.50\\ 
  \hline
    \multirow{2}{*}{$\mathcal{S}_0$ \{111\}} & LA &  21.88 & 22.57\\ 
    			        &TA &  11.01 & 11.20\\
  \hline
 \multicolumn{4}{|c|}{Low symmetry phase}  \\ 
   \hline
   $\mathcal{S}_2$ & LA & 24.83 & 25.37\\%(24.43 + TA 8.74)\\
   \hline
   \multirow{3}{*}{$\mathcal{S}_1$} & LA &  24.77 & 23.34\\ 
    			        &TA$_{1}$ &  10.72 & 13.79\\%no signal(12.2)\\
    			         &TA$_{2}$ &  7.86 & 12.10\\%(\~10)\\
  \hline
  \multirow{3}{*}{$\mathcal{S}_0$} & LA &  22.22 & 22.80\\ 
    			        &TA$_{1}$ &  12.15 & 12.17\\
    			          &TA$_{2}$ &  10.86 & 10.33\\
\hline
\end{tabular}
\caption{Phonon frequencies evaluated from peak positions in BLS spectra under excitation along different crystallographic directions at room temperature 293~K in cubic phase and at 5~K in tetragonal phase under 542~nm laser excitation.}
\label{table:PhFreq}
\end{table*}

The spectral position of the Brillouin peaks can be directly related to the components of the elastic stiffness tensor, see details in Appendix~\ref{sec:c_to_v}. In particular, $c_{11}$ may be directly calculated from the frequency of LA phonon measured at \{001\} surface while its change for other surfaces is defined by the $c_{44}-(c_{11}-c_{12})/2$ characterizing the deviation of elastic properties of cubic material from the isotropic one. From the experimental data, we conclude that both materials can be treated as isotropic with a good precision, which is beyond actual experimental resolution. The exact numbers will be discussed below.

The BLS data along different crystallographic directions provide full information about the elastic properties of the material. From the results shown in Fig.~\ref{fig:rtbls} we determine the elastic tensors of the perovskite crystals under study. 
For the cubic phase, all three components of the elastic tensor are calculated directly from the peak positions. In particular, the position of single LA phonon peak at \{001\} surface is recalculated to $c_{11}$ as:
\begin{equation}
  c_{11} = \frac{\rho \lambda^2}{4 n_r^2} \left( \delta\nu_{LA,001} \right)^2\,.
\end{equation}
Other components may be calculated analogously. Note, that in the cubic phase TA phonon modes contribute to the Brillouin signal only at \{111\} surface. Here, we use the following procedure. We calculate all three constants from the signal at \{001\} and \{111\} surfaces and estimate the experimental error by comparing signals at different wavelengths and \{110\} surface. 

For Cs\tsub{2}AgBiBr\tsub{6} we obtain $ c_{11} = 40.2\pm0.3$~GPa, $c_{12} = 19.4\pm0.3$~GPa, $c_{44} = 10.1\pm0.3$~GPa. Here, the quoted $\pm$ values reflect the estimated accuracy of the extracted constants, inferred from the fact that the peak from \{110\} facet is shifted by about 0.2~GHz from its expected position. These results are in reasonable agreement with experimental data obtained by  nanoindentation and inelastic X-ray scattering in Ref.~\cite{Lun2022} and density functional theory calculations (DFT) in Ref.~\cite{Horiachyi2025}, see Table~\ref{tab:elastic_constants}. Using the same procedure we evaluate the following constants for Cs\tsub{2}AgBiCl\tsub{6}: $c_{11} = 42.8\pm1$~GPa, $c_{12} = 21.2\pm1$~GPa, $c_{44} = 8.55\pm1$~GPa. Although the uncertainty is somewhat larger in this case, it remains within the resolution of the setup, corresponding to a peak-position accuracy of about $\pm0.7$~GHz.

\begin{table}
\centering
\caption{Elastic constants for Cs$_2$AgBiBr$_6$ and Cs$_2$AgBiCl$_6$ from different sources and extracted from current BLS data. All values are given in GPa.
}
\label{tab:elastic_constants}
\begin{tabular}{|l|c|ccc|}
\hline
& Reference & $C_{11}$  &  $C_{44}$ & $C_{12}$\\
\hline
\textbf{Cs$_2$AgBiBr$_6$:} & & & &\\
 Ref.~\cite{Lun2022} & Experiment  & 44.5 & 8.6  & 17.6 \\
 Ref.~\cite{Horiachyi2025} & Theory 
                      & 45.7 & 7.0 & 17.2 \\
 Ref.~\cite{BLS-2025} & Theory & 33.6 & 7.25 & 14.5 \\
this work   & Experiment   & 40.2 & 10.1 & 19.4 \\ \hline
\textbf{Cs$_2$AgBiCl$_6$:} &&&& \\
 Ref.~\cite{BLS-2025} & Theory & 42.1 & 9.01 & 15.1 \\
 this work & Experiment & 42.8 & 8.55 & 21.2 \\
\hline
%\label{table:comp}
\end{tabular}
\end{table}

\subsection{Tetragonal phase at low temperature}
\label{sec:lowT}

At cryogenic temperatures, Cs\tsub{2}AgBiBr\tsub{6} is known to have low symmetry crystallographic phase. Below 120~K, it is in the tetragonal phase with \#87 space group \cite{Schade2019}. There are recent reports that at a temperature of 39~K another structural transition from tetragonal to monoclinic or triclinic phase takes place~\cite{Cohen2022,Tower2025}. The latter is not accompanied by a significant change of acoustic properties~\cite{Horiachyi2025} and below we will assume that the material has tetragonal phase. The crystallographic directions after the cubic-to-tetragonal phase transition are ambiguous and extra work needed to identify them. In cubic phase we identified the surfaces normal to the corresponding crystallographic directions: $\left\langle 001 \right\rangle$, $\left\langle 110 \right\rangle$ and $\left\langle 111 \right\rangle$. To avoid confusion, for these surfaces in tetragonal phase we will use correspondingly $\mathcal{S}_2$, $\mathcal{S}_1$, $\mathcal{S}_0$.
\begin{figure}
\centering
\includegraphics[width=.99\linewidth]{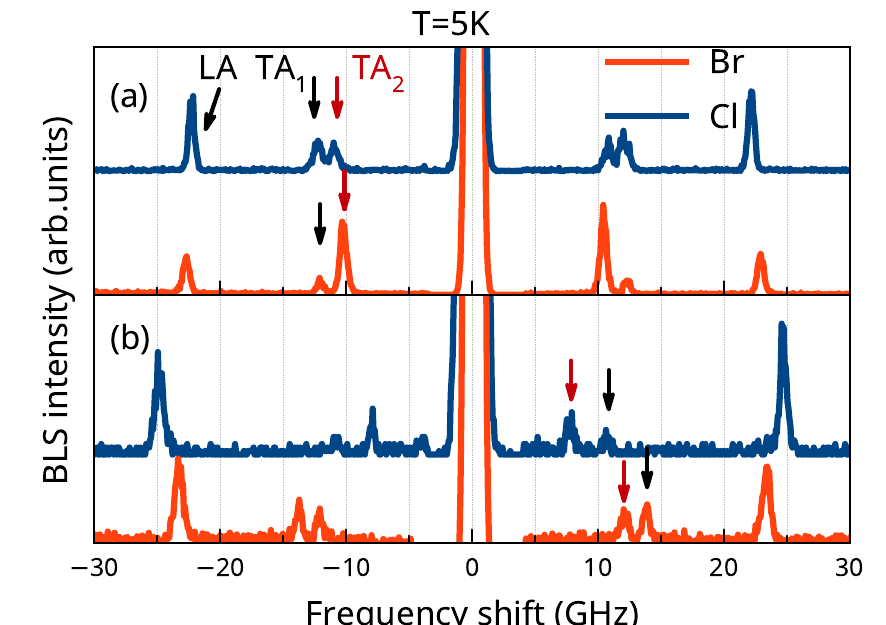}
  \caption{Low temperature BLS spectra at $T\approx 5$~K. BLS spectra under excitation with 542~nm laser for Cs$_{2}$AgBiCl$_{6}$ (blue) and Cs$_{2}$AgBiBr$_{6}$ (red) measured at 5~K for $\mathcal{S}_0$ (a) and $\mathcal{S}_1$ (b) surfaces.}
  \label{fig:5kbls}
\end{figure}

Figure~\ref{fig:5kbls} summarizes the BLS data obtained at low temperature $T=5$~K for two different facets. The spectrum measured at $\mathcal{S}_0$ surface of Cs$_{2}$AgBiCl$_{6}$ (see Fig.\ref{fig:5kbls}(a)) show three peaks with maxima at $22.22$~GHz, $12.15$~GHz and $10.86$~GHz, that are attributed to one longitudinal phonon and two transverse phonons (TA$_1$ and TA$_2$), respectively. For Cs$_{2}$AgBiBr$_{6}$ under the same conditions we observe LA, TA$_1$, and TA$_2$ peaks with a frequencies of $22.80$~GHz, $12.17$~GHz and $10.33$~GHz, respectively. 

The presence of three different peaks measured at 5~K at  $\mathcal{S}_0$ surface in case of Cs$_{2}$AgBiBr$_{6}$ is attributed to the transition to a lower symmetry crystal phase. In tetragonal phase the degeneracy of shear acoustic phonons is lifted and we can observe two transverse acoustic phonons with different frequencies. To the best of our knowledge, there is no information about structural phase transitions for Cs$_{2}$AgBiCl$_{6}$. Splitting of transverse acoustic phonons means that the symmetry of the crystal is lowered compared to the one at room temperature. By analogy with other halide perovskite crystals, we suggest that the chloride crystal undergoes a structural phase transition. Note, that the alignment of $c$-axis along one of the three possible directions in tetragonal phase is not important for the evaluation of peak positions measured at $\mathcal{S}_0$ surface.

Spectra obtained at 5~K at the $\mathcal{S}_1$ surface also show three different peaks in both samples, compared to one peak measured at room temperature. For Cs$_{2}$AgBiCl$_{6}$ peak frequencies are $24.77$~GHz, $10.72$~GHz and $7.86$~GHz, while for Cs$_{2}$AgBiCl$_{6}$ the corresponding frequencies are $23.34$~GHz,  $13.79$~GHz and $12.10$~GHz.  Here, we observe very strong variation of phonon frequencies, particularly for TA modes, when comparing bromide and chloride samples. Such a big difference in phonon frequencies could be attributed to the fact that in a low symmetry crystal the directions of the cubic phase normal to $\mathcal{S}_1$ surface are no longer equivalent. Therefore, the elastic properties depend on the alignment of $c$-axis in low symmetry phase and can differ significantly  along these directions. 

In other words, spectra presented in Fig.~\ref{fig:5kbls}b are likely obtained from crystal surfaces that are equivalent in cubic phases and not equivalent in tetragonal phase.
 
In contrast to the cubic phase, the amount of BLS data in tetragonal phase are not enough to extract all elastic constants, since their number increases from three to seven, especially taking into account the experimental uncertainties in the peaks positions. To compare the values of elastic constants in low symmetry phase one requires larger series of measurements along other crystallographic directions.
However, as observed for the cubic phase of both materials, similar peak positions obtained along $\langle 111 \rangle$ direction correspond to close values of the elastic tensor components and we expect to see similar values for elastic constants in low symmetry crystallographic phase.

\subsection{Temperature dependence}
\label{sec:Tdep}

\begin{figure*}
\centering
\includegraphics[width=.99\linewidth]{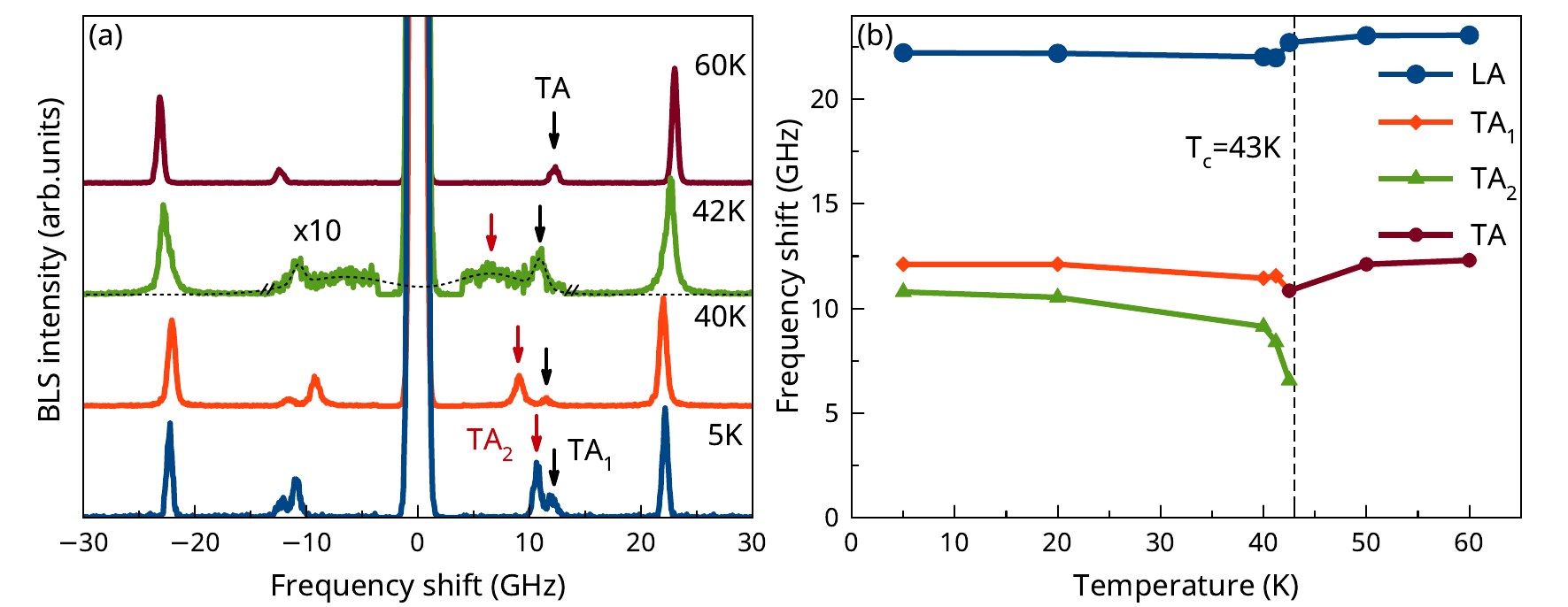}
  \caption{Temperature dependence of the phonon frequencies in Cs$_{2}$AgBiCl$_{6}$ along $\langle 111 \rangle$ crystallographic direction  under excitation with 542~nm laser. (a)  BLS spectra measured at different temperatures. (b) BLS peak positions as a function of temperature. Blue points correspond to LA phonons, red - TA$_1$ phonons, green - TA$_2$ phonons. }
  \label{fig:tempbls}
\end{figure*}

In order to determine the temperature of structural phase transition phase $T_C$ in Cs$_{2}$AgBiCl$_{6}$ we study the temperature dependence of BLS spectra for $\mathcal{S}_0$ surface when scattering on all acoustic phonons is well pronounced and positions of the peaks do not depend on the particular alignment of $c$-axis. The results are summarized in Fig.~\ref{fig:tempbls}. The spectra at selected temperatures are shown in Fig.~\ref{fig:tempbls}(a). We observe that an increase in temperature leads to a shift of the TA phonon peaks toward lower frequencies. This effect is particularly pronounced for the TA$_2$ peak, which becomes very weak and broad at a temperature of 42~K (see green curve with the TA$_1$ and TA$_2$ peaks labeled by black and red arrows, respectively). A further increase in the temperature leads to transformation of two TA peaks in to a single TA peak at a higher frequency. Above 43~K the BLS spectra comprise only two peaks with frequencies corresponding to LA and TA phonons as observed at room temperature in Fig.~\ref{fig:rtbls}(b). Therefore, we conclude that in Cs$_{2}$AgBiCl$_{6}$ the structural phase transition between low temperature tetragonal and high temperature cubic phases takes place at $T_C=43$~K. 

The temperature dependence of the peak positions is shown in Fig.~\ref{fig:tempbls}(b). Maximum of LA phonon in Cs$_{2}$AgBiCl$_{6}$ shifts by about $0.8$~GHz from $22.22$~GHz before the phase transition to $23.07$ after the phase transition, while TA$_1$ shifts from $12.11$~GHz at 5~K to $12.38$ in a cubic phase at 60~K. TA$_2$ frequency changes from $10.79$~GHz to up to $6.50$~GHz at $42~$K. We emphasize that close to $T_C$ in tetragonal phase we observe a very strong frequency shift of TA$_2$ mode to lower energies with increasing the temperature. The softening of transverse acoustic modes across the tetragonal–cubic transition can be attributed to an enhanced interaction between acoustic and optical phonons. A closely related behavior is well established for SrTiO$_3$, where the cubic–tetragonal antiferrodistortive transition is accompanied by pronounced acoustic–optical phonon mixing~\cite{Fleury1968, HehlenTagantsev1998}.  We note that the temperature dependence of LA, TA$_1$ and TA$_2$ peak positions is very similar to that observed for Cs$_{2}$AgBiBr$_{6}$ as reported in Ref.~\cite{Horiachyi2025}. However, the temperature of phase transition is lowered from 122~K to about 43~K. It is worth to mention that local heating of the sample in the excitation area may lead to local increase of temperature. Therefore, we keep this value as a lower estimate, meaning that the real $T_C$ could be several degrees larger than 43~K.

%%%%% Discussion %%%%%%%%%%%%%%%%%%%%%%%%%%%%%%%
\section{Discussion}
\label{sec:discuss}

A complete set of experimentally measured elastic properties is available for the lead-free double perovskite Cs$_{2}$AgBiBr$_{6}$ in the cubic phase, where the elastic constants were determined using nanoindentation and inelastic X-ray scattering~\cite{Lun2022}. To our knowledge, no experimental data on the elastic constants of Cs$_2$AgBiCl$_6$ have been reported. For this crystal, the BLS spectra are available only along the $\langle111\rangle$ crystallographic direction, which is insufficient to determine the complete set of elastic constants~\cite{BLS-2025}. Moreover, the sound velocities along $\langle111\rangle$ were obtained using refractive indices derived from DFT calculations. In the present work, by evaluating and comparing BLS data obtained from three different crystal facets, we were able to determine the full set of elastic constants in the cubic phase (see Table~\ref{tab:elastic_constants}). We emphasize that these values are very similar for both the bromide and chloride compounds in the cubic phase.

Next, we compare the elastic constants listed in Table~\ref{tab:elastic_constants} with previously reported data for organic lead halide perovskites obtained using inelastic neutron scattering and BLS in Ref.~\cite{Ferreira-2018}. First, similarly to lead halides the value for $c_{11}$ decreases with the increase of crystal lattice constant, i.e. it is larger for bromide compared to chloride double perovskite. Second, in lead-halides the longitudinal/transverse ratio $c_{11}/c_{44}$ is large (about 10), whereas in double perovskites it is significantly smaller, being about 4. Third, the Zener anisotropy parameter $A=2c_{44}/(c_{11}-c_{12})$ equals to 0.97 and 0.79 in Cs$_{2}$AgBiBr$_{6}$ and Cs$_{2}$AgBiCl$_{6}$, respectively. This anisotropy is substantially smaller than in lead halide perovskites, where $A$ deviates from unity in the range 0.3–0.5 as reported in Ref.~\cite{Ferreira-2018}.  

From analysis of temperature dependent BLS data, we conclude that similarly to Cs$_{2}$AgBiBr$_{6}$, the chloride perovskite shows a clear structural phase transition accompanied by the softening of the low frequency TA mode. Compared to  Cs$_{2}$AgBiBr$_{6}$, the phase transition in Cs$_{2}$AgBiCl$_{6}$ occurs at lower temperature. Similar behavior was reported for MAPBr$_{3}$ and MAPCl$_{3}$~\cite{Simenas-2024}, where structural phase transition in perovskite with Br takes place at higher temperature due to larger stiffness of this material.  

%%%%% Conclusion %%%%%%%%%%%%%%%%%%%%%%%%%%%

\section{Conclusion}
\label{sec:conclusion}

We experimentally determine the full set of elastic constants in Cs$_2$AgBiBr$_6$ and Cs$_2$AgBiCl$_6$ in cubic phase. In Cs$_2$AgBiCl$_6$, a phase transition from the cubic phase to a lower-symmetry crystal structure is observed at 43~K, manifested by the splitting of transverse acoustic modes and a pronounced softening of a low-frequency mode. 
Double perovskites demonstrate smaller longitudinal/transverse ratio and large degree of isotropy as compared with lead halide perovskites.
Further measurements of multiple facets in combination with polarization resolved studies would allow access to a larger set of elastic constants in the low-temperature phase and enable a more detailed characterization of the elastic anisotropy.

%%%%% Acknowledgments %%%%%%%%%%%%%%%%%%%%%%%%%%%

\section*{Acknowledgments}
The Dortmund and Würzburg groups acknowledge financial support from the Deutsche Forschungsgemeinschaft within the SPP 2196 (project no. 506623857). D.O.H. acknowledges the Deutsche Forschungsgemeinschaft (project no. 536987509). 

\appendix

\section{Density}
From Ref.~\cite{Schade2019}, the volume for Cs\tsub{2}AgBiBr\tsub{6} one may conclude that the volume is linear with temperature, the volume per functional unit is $\approx 351+T/40$~\AA$^{3}$. The mass of the functional unit is 1062.094~a.u. ($=1.7636\cdot 10^{-24}$~kg). As a result, the density is $\rho_{\rm Br}(T) \approx 5024.5 - 0.35T$~kg/m$^3$.

Assuming that the lattice constant ratio in Cs\tsub{2}AgBiCl\tsub{6} and Cs\tsub{2}AgBiBr\tsub{6} does not depend on temperaure, we take the values from XRD measurements and using the mass of functional unit 795.37~a.u ($1.3207\cdot 10^{-24}$~kg), one obtains for the density $\rho_{\rm Cl}(T) \approx 4303.8 - 0.3 T$~kg/m$^3$.

\section{Acoustic phonons velocities}\label{sec:c_to_v}
The velocities of the acoustic phonon modes $V_s$ are the eigenvalues of the equation
\begin{equation}\label{eq:eigen_ph}
  \left( \rho V_s^2 \delta_{il} - c_{ijkl} Q_j Q_k \right) U_l =0\,,
\end{equation}
where $\rho$ is the density of material, $c_{ijkl}$ are elastic constants of the material and ${\bf Q}$ is the direction of the phonon wave vector. 

In cubic material, there are three independent components of the elastic tensor denoted as $c_{11}$, $c_{12}$, $c_{44}$. For the analysis, it is more convenient to use $c_{11}$ and $c_{44}$ which are directly connected with sound velocities of isotropic material and $c_{\delta \rm c} = c_{44}-(c_{11}-c_{12})/2$ which characterizes the anisotropy of the material and related with  Zener anisotropy index \cite{Zener} $A = 2c_{44}/(c_{11}-c_{12})$ as $c_{\delta \rm c} = c_{44}(1-1/A) $. In isotropic material, $A=1$ while $c_{\delta c}=0$.
With these notations, from Eq.~\eqref{eq:eigen_ph} it is easy to find (c.f. Table XIII in Ref.~\cite{Vacher1972}) velocities for the wave vector along high symmetry directions:
\begin{align}
  {\bf Q} &\| [001]: &  \rho V_{TA}^2  & = c_{44}\,, & \rho V_{LA}^2 & = c_{11}\,;\\
  {\bf Q} &\| [110]: &  \rho V_{TA1}^2 & = c_{44}\,, & \rho V_{TA2}^2 & = c_{44}-c_{\delta c}  \,,\\\nonumber
  & &&  &               \rho V_{LA}^2  & = c_{11}+c_{\delta c}\,; \\ 
  {\bf Q} &\| [111]: & &&  \rho V_{TA}^2  & = c_{44}-\frac23 c_{\delta c}\,, \\
  & &&  & \rho V_{LA}^2 & = c_{11} + \frac43 c_{\delta c}\,.  \nonumber
\end{align}

%%%  REFERENCES  %%%%%%%%%%%%%%%%%%%%%%%%%%%%%%%%
\bibliographystyle{unsrt}
\bibliography{refs}

@book{Zener,
  title={Elasticity and Anelasticity of Metals},
  author={Zener, C.},
  isbn={9780226980546},
  lccn={lc48003902},
  series={Chicago University Committee on Publications in the Physical Sciences},
  url={https://books.google.de/books?id=FKcZAAAAIAAJ},
  year={1948},
  publisher={University of Chicago Press}
}

@misc{Tower2025,
      title={Low-temperature structural instabilities of the halide double perovskite {Cs$_2$AgBiBr$_6$} investigated via x-ray diffraction and infrared phonons}, 
      author={Collin Tower and Fereidoon S. Razavi and Jeremy Dion and Jürgen Nuss and Reinhard K. Kremer and Maureen Reedyk},
      year={2025},
      eprint={2505.10563},
      archivePrefix={arXiv},
      primaryClass={cond-mat.mtrl-sci},
      url={https://arxiv.org/abs/2505.10563}, 
}

@article{Horiachyi2025,
  author = {
    D. O. Horiachyi and M. O. Nestoklon and I. A. Akimov and A. V. Trifonov and 
    N. V. Siverin and N. E. Kopteva and A. N. Kosarev and D. R. Yakovlev and 
    V. E. Gusev and M. Fries and  O. Trukhina and V. Dyakonov and Manfred Bayer},
  title = {Efficient launching of shear phonons in photostrictive halide perovskites},
  journal = {Science Advances},
  year = {2025},
  volume = {11},
  pages = {eadw9172},
  doi = {10.1126/sciadv.adw9172}
}

@Article{Lun2022,
  author={Lun, Y.
  and Liu, J.
  and Wei, B.
  and Gao, Z.
  and Wang, X.
  and Hong, J.},
  title={Elastic Properties of Photovoltaic Single Crystal {Cs2AgBiBr6}},
  journal={Experimental Mechanics},
  year={2022},
  month={Jan},
  day={01},
  volume={62},
  number={1},
  pages={117-123},
  issn={1741-2765},
  doi={10.1007/s11340-021-00768-9},
  url={https://doi.org/10.1007/s11340-021-00768-9}
}

@article{Vacher1972,
  title = {Brillouin Scattering: A Tool for the Measurement of Elastic and Photoelastic Constants},
  author = {Vacher, R. and Boyer, L.},
  journal = {Phys. Rev. B},
  volume = {6},
  issue = {2},
  pages = {639--673},
  numpages = {0},
  year = {1972},
  month = {Jul},
  publisher = {American Physical Society},
  doi = {10.1103/PhysRevB.6.639},
  url = {https://link.aps.org/doi/10.1103/PhysRevB.6.639}
}

@Article{Polyanskiy2024,
author={Polyanskiy, Mikhail N.},
title={Refractiveindex.info database of optical constants},
journal={Scientific Data},
year={2024},
month={Jan},
day={18},
volume={11},
number={1},
pages={94},
issn={2052-4463},
doi={10.1038/s41597-023-02898-2},
url={https://doi.org/10.1038/s41597-023-02898-2}
}

@Article{Gray2019,
  author ="Gray, Matthew B. and McClure, Eric T. and Woodward, Patrick M.",
  title  ="{Cs2AgBiBr6-xClx} solid solutions -- band gap engineering with halide double perovskites",
  journal  ="J. Mater. Chem. C",
  year  ="2019",
  volume  ="7",
  issue  ="31",
  pages  ="9686-9689",
  publisher  ="The Royal Society of Chemistry",
  doi  ="10.1039/C9TC02674F",
}

@article{Schade2019,
  title = {Structural and Optical Properties of {C}s$_{2}${A}g{B}i{B}r$_{6}$ Double Perovskite},
  author = {Schade, L. and  Wright, A. D. and Johnson, R. D. and  Dollmann, M. and Wenger, B. and Nayak, P. K. and Prabhakaran, D. and  Herz, L.M. and Nicholas, R. and Snaith, H. J. and Radaelli, P. G.},
  journal = {ACS Energy Lett},
  volume = {4},
  pages = {299},
  numpages = {6},
  year = {2019},
  doi = {10.1021/acsenergylett.8b02090},
}

@article{McClure-2016,
author = {McClure, Eric T. and Ball, Molly R. and Windl, Wolfgang and Woodward, Patrick M.},
title = {ChemInform Abstract: {Cs2AgBiX6} ({X}: {Br}, {Cl}): New Visible Light Absorbing, Lead-Free Halide Perovskite Semiconductors.},
journal = {ChemInform},
volume = {47},
number = {20},
pages = {},
doi = {https://doi.org/10.1002/chin.201620017},
url = {https://onlinelibrary.wiley.com/doi/abs/10.1002/chin.201620017},
eprint = {https://onlinelibrary.wiley.com/doi/pdf/10.1002/chin.201620017},
year = {2016}
}

@article{Jain-2025,
author = {Jain, Pulkita and Tran, Minh N. and Cleveland, Iver J. and Liu, Yukun and Sarp, Seda and Aydil, Eray S.},
title = {Vapor Deposition and Optical Properties of {Cs2AgBiCl6} Thin Films},
journal = {The Journal of Physical Chemistry C},
volume = {129},
number = {11},
pages = {5301-5311},
year = {2025},
doi = {10.1021/acs.jpcc.4c06622},
URL = {https://doi.org/10.1021/acs.jpcc.4c06622},
eprint = {https://doi.org/10.1021/acs.jpcc.4c06622}
}

@Article{Zelewski-2019,
author ={Zelewski, S. J. and Urban, J. M. and Surrente, A. and Maude, D. K. and Kuc, A. and Schade, L. and Johnson, R. D. and Dollmann, M. and Nayak, P. K. and Snaith, H. J. and Radaelli, P. and Kudrawiec, R. and Nicholas, R. J. and Plochocka, P. and Baranowski, M.},
title  ={Revealing the nature of photoluminescence emission in the metal-halide double perovskite {Cs2AgBiBr6}},
journal  = {J. Mater. Chem. C},
year  = {2019},
volume  = {7},
issue  = {27},
pages  = {8350-8356},
publisher  = {The Royal Society of Chemistry},
doi  = {10.1039/C9TC02402F},
url  = {http://dx.doi.org/10.1039/C9TC02402F},
}

@article{Ferreira-2018,
  title = {Elastic Softness of Hybrid Lead Halide Perovskites},
  author = {Ferreira, A. C. and L\'etoublon, A. and Paofai, S. and Raymond, S. and Ecolivet, C. and Ruffl\'e, B. and Cordier, S. and Katan, C. and Saidaminov, M. I. and Zhumekenov, A. A. and Bakr, O. M. and Even, J. and Bourges, P.},
  journal = {Phys. Rev. Lett.},
  volume = {121},
  issue = {8},
  pages = {085502},
  numpages = {6},
  year = {2018},
  month = {Aug},
  publisher = {American Physical Society},
  doi = {10.1103/PhysRevLett.121.085502},
  url = {https://link.aps.org/doi/10.1103/PhysRevLett.121.085502}
}

@article{BLS-2025,
title = {Vibrational, optical and elastic properties of {Cs2AgBiX6} ({X}= {Br}, {Cl}) double perovskite materials: {DFT}, {PL}, {R}aman and {B}rillouin scattering studies},
journal = {Journal of Alloys and Compounds},
volume = {1010},
pages = {176968},
year = {2025},
issn = {0925-8388},
doi = {https://doi.org/10.1016/j.jallcom.2024.176968},
url = {https://www.sciencedirect.com/science/article/pii/S0925838824035552},
author = {Furqanul Hassan Naqvi and Jae-Hyeon Ko and Tae Heon Kim and Chang Won Ahn and Younghun Hwang},
}

@article{BLS-2022,
author = {Pang, Simin and Liu, Xinbao and Luo, Jiajun and Xie, Yaru and Tang, Jiang and Meng, Sheng and Tan, Ping-Heng and Zhang, Jun},
title = {Brillouin Light Scattering of Halide Double Perovskite},
journal = {Advanced Photonics Research},
volume = {3},
number = {8},
pages = {2100222},
doi = {https://doi.org/10.1002/adpr.202100222},
url = {https://advanced.onlinelibrary.wiley.com/doi/abs/10.1002/adpr.202100222},
eprint = {https://advanced.onlinelibrary.wiley.com/doi/pdf/10.1002/adpr.202100222},
year = {2022}
}

@article{Simenas-2024,
author = {Simenas, Mantas and Gagor, Anna and Banys, Juras and Maczka, Miroslaw},
title = {Phase Transitions and Dynamics in Mixed Three- and Low-Dimensional Lead Halide Perovskites},
journal = {Chemical Reviews},
volume = {124},
number = {5},
pages = {2281-2326},
year = {2024},
doi = {10.1021/acs.chemrev.3c00532},
    note ={PMID: 38421808},
URL = { https://doi.org/10.1021/acs.chemrev.3c00532},
eprint = {https://doi.org/10.1021/acs.chemrev.3c00532}
}

@article{Cohen2022,
author = {Cohen, Adi and Brenner, Thomas M. and Klarbring, Johan and Sharma, Rituraj and Fabini, Douglas H. and Korobko, Roman and Nayak, Pabitra K. and Hellman, Olle and Yaffe, Omer},
title = {Diverging Expressions of Anharmonicity in Halide Perovskites},
journal = {Advanced Materials},
volume = {34},
number = {14},
pages = {2107932},
doi = {https://doi.org/10.1002/adma.202107932},
url = {https://advanced.onlinelibrary.wiley.com/doi/abs/10.1002/adma.202107932},
year = {2022}
}

@article{Fleury1968,
  title = {Soft Phonon Modes and the 110\ifmmode^\circ\else\textdegree\fi{}{K} Phase Transition in {SrTi${\mathrm{O}}_{3}$}},
  author = {Fleury, P. A. and Scott, J. F. and Worlock, J. M.},
  journal = {Phys. Rev. Lett.},
  volume = {21},
  issue = {1},
  pages = {16--19},
  numpages = {0},
  year = {1968},
  month = {Jul},
  publisher = {American Physical Society},
  doi = {10.1103/PhysRevLett.21.16},
  url = {https://link.aps.org/doi/10.1103/PhysRevLett.21.16}
}

@article{HehlenTagantsev1998,
  title = {Brillouin-scattering observation of the TA-TO coupling in ${\mathrm{SrTiO}}_{3}$},
  author = {Hehlen, B. and Arzel, L. and Tagantsev, A. K. and Courtens, E. and Inaba, Y. and Yamanaka, A. and Inoue, K.},
  journal = {Phys. Rev. B},
  volume = {57},
  issue = {22},
  pages = {R13989--R13992},
  numpages = {0},
  year = {1998},
  month = {Jun},
  publisher = {American Physical Society},
  doi = {10.1103/PhysRevB.57.R13989},
  url = {https://link.aps.org/doi/10.1103/PhysRevB.57.R13989}
}

@article{Jia2025,
  title = {},
  author = {Jia, Zhenrong and Guo, Xiao and Yin, Xinxing and Sun, Ming and Qiao, Jiawei and Jiang, Xinyu and Wang, Xi and and Wang, Yuduan and Dong, Zijing and Shi, Zhuojie and Kuan, Chun-Hsiao and Hu, Jingcong and  Zhou, Qilin and Jia, Xiangkun and Chen, Jinxi and  Wei, Zhouyin and Liu, Shunchang and Liang, Haoming and Li, Nengxu and Lee, Ling Kai and Guo, Renjun and Roth, Stephan V. and M\"uller-Buschbaum, Peter and Hao, Xiaotao and Du, Xiaoyan and Hou, Yi},
  journal = {Nature},
  volume = {643},
  pages = {104-110},
  year = {2025},
  doi = {10.1038/s41586-025-09181-x},
  url = {https://doi.org/10.1038/s41586-025-09181-x}
}

@article{Dyakonov2021,
    author = {Schmidt-Mende, Lukas and Dyakonov, Vladimir and Olthof, Selina and Ünlü, Feray and Lê, Khan Moritz Trong and Mathur, Sanjay and Karabanov, Andrei D. and Lupascu, Doru C. and Herz, Laura M. and Hinderhofer, Alexander and Schreiber, Frank and Chernikov, Alexey and Egger, David A. and Shargaieva, Oleksandra and Cocchi, Caterina and Unger, Eva and Saliba, Michael and Byranvand, Mahdi Malekshahi and Kroll, Martin and Nehm, Frederik and Leo, Karl and Redinger, Alex and Höcker, Julian and Kirchartz, Thomas and Warby, Jonathan and Gutierrez-Partida, Emilio and Neher, Dieter and Stolterfoht, Martin and Würfel, Uli and Unmüssig, Moritz and Herterich, Jan and Baretzky, Clemens and Mohanraj, John and Thelakkat, Mukundan and Maheu, Clément and Jaegermann, Wolfram and Mayer, Thomas and Rieger, Janek and Fauster, Thomas and Niesner, Daniel and Yang, Fengjiu and Albrecht, Steve and Riedl, Thomas and Fakharuddin, Azhar and Vasilopoulou, Maria and Vaynzof, Yana and Moia, Davide and Maier, Joachim and Franckevičius, Marius and Gulbinas, Vidmantas and Kerner, Ross A. and Zhao, Lianfeng and Rand, Barry P. and Glück, Nadja and Bein, Thomas and Matteocci, Fabio and Castriotta, Luigi Angelo and Di Carlo, Aldo and Scheffler, Matthias and Draxl, Claudia},
    title = {Roadmap on organic–inorganic hybrid perovskite semiconductors and devices},
    journal = {APL Materials},
    volume = {9},
    number = {10},
    pages = {109202},
    year = {2021},
    month = {10},
    issn = {2166-532X},
    doi = {10.1063/5.0047616},
    url = {https://doi.org/10.1063/5.0047616},
    eprint = {https://pubs.aip.org/aip/apm/article-pdf/doi/10.1063/5.0047616/19735731/109202_1_online.pdf},
}

@article{Tress2022,
author = {Tress, Wolfgang and Sirtl, Maximilian T.},
title = {{Cs$_2$AgBiBr$_6$} Double Perovskites as Lead-Free Alternatives for Perovskite Solar Cells?},
journal = {Solar RRL},
volume = {6},
number = {2},
pages = {2100770},
doi = {https://doi.org/10.1002/solr.202100770},
url = {https://onlinelibrary.wiley.com/doi/abs/10.1002/solr.202100770},
eprint = {https://onlinelibrary.wiley.com/doi/pdf/10.1002/solr.202100770},
year = {2022}
}

@article{Zhang2022,
author = {Zhang, Zeyu and Sun, Qingde and Lu, Yue and Lu, Feng and Mu, Xulin and Wei, Su-Huai and Sui, Manling},
title = {Hydrogenated {Cs$_2$AgBiBr$_6$} for significantly improved efficiency of lead-free inorganic double perovskite solar cell},
journal = {Nature Communications},
volume = {13},
pages = {3397},
doi = {10.1038/s41467-022-31016-w},
url = {https://doi.org/10.1038/s41467-022-31016-w},
year = {2022}
}

@article{Yamada2022,
author = {Yamada, Yasuhiro and Kanemitsu, Yoshihiko},
title = {Electron-phonon interactions in halide perovskites},
journal = {NPG Asia Materials},
volume = {14},
pages = {48},
doi = {10.1038/s41427-022-00394-4},
url = {https://doi.org/10.1038/s41427-022-00394-4},
year = {2022}
}

@article{Maris1986,
  title = {Surface generation and detection of phonons by picosecond light pulses},
  author = {Thomsen, C. and Grahn, H. T. and Maris, H. J. and Tauc, J.},
  journal = {Phys. Rev. B},
  volume = {34},
  issue = {6},
  pages = {4129--4138},
  numpages = {0},
  year = {1986},
  month = {Sep},
  publisher = {American Physical Society},
  doi = {10.1103/PhysRevB.34.4129},
  url = {https://link.aps.org/doi/10.1103/PhysRevB.34.4129}
}

@article{Young2012,
  title = {Picosecond strain pulses generated by a supersonically expanding electron-hole plasma in GaAs},
  author = {Young, E. S. K. and Akimov, A. V. and Campion, R. P. and Kent, A. J. and Gusev, V.},
  journal = {Phys. Rev. B},
  volume = {86},
  issue = {15},
  pages = {155207},
  numpages = {13},
  year = {2012},
  month = {Oct},
  publisher = {American Physical Society},
  doi = {10.1103/PhysRevB.86.155207},
  url = {https://link.aps.org/doi/10.1103/PhysRevB.86.155207}
}

@article{Nelson1994,
  title = {Femtosecond time-resolved spectroscopy of soft modes in structural phase transitions of perovskites},
  author = {Dougherty, Thomas P. and Wiederrecht, Gary P. and Nelson, Keith A. and Garrett, Mark H. and Jenssen, Hans P. and Warde, Cardinal},
  journal = {Phys. Rev. B},
  volume = {50},
  issue = {13},
  pages = {8996--9019},
  numpages = {0},
  year = {1994},
  month = {Oct},
  publisher = {American Physical Society},
  doi = {10.1103/PhysRevB.50.8996},
  url = {https://link.aps.org/doi/10.1103/PhysRevB.50.8996}
}

@article{Ruello2015,
title = {Physical mechanisms of coherent acoustic phonons generation by ultrafast laser action},
journal = {Ultrasonics},
volume = {56},
pages = {21-35},
year = {2015},
issn = {0041-624X},
doi = {https://doi.org/10.1016/j.ultras.2014.06.004},
url = {https://www.sciencedirect.com/science/article/pii/S0041624X1400153X},
author = {Pascal Ruello and Vitalyi E. Gusev},
}

@article{Steele2020,
author = {Steele, Julian A. and Lai, Minliang and Zhang, Ye and Lin, Zhenni and Hofkens, Johan and Roeffaers, Maarten B. J. and Yang, Peidong},
title = {Phase Transitions and Anion Exchange in All-Inorganic Halide Perovskites},
journal = {Accounts of Materials Research},
volume = {1},
number = {1},
pages = {3-15},
year = {2020},
doi = {10.1021/accountsmr.0c00009},
URL = {https://doi.org/10.1021/accountsmr.0c00009},
eprint = {https://doi.org/10.1021/accountsmr.0c00009}
}

@article{Kirstein2020,
author = { Kirstein, E. and Yakovlev, D. R. and Glazov, M. M. and Zhukov, E. A. and Kudlacik, D. and Kalitukha, I. V. and Sapega, V. F. and Dimitriev, G. S. and  Semina, M. A. and  Nestoklon, M. O. and Ivchenko, E. L. and Kopteva, N. E. and Dirin, D. N. and Nazarenko, O.
 Kovalenko, M. V. and  Baumann, A. and H\"ocker, J. and  Dyakonov, V. and Bayer, M.},
title = {},
journal = {Nature Communications},
volume = {13},
pages = {3062},
year = {2022},
doi = {10.1038/s41467-022-30701-0},
URL = {https://doi.org/10.1038/s41467-022-30701-0},
}

@article{Shaller2017,
author = {Guo, Peijun and Xia, Yi and Gong, Jue and Stoumpos, Constantinos C. and McCall, Kyle M. and Alexander, Grant C. B. and Ma, Zhiyuan and Zhou, Hua and Gosztola, David J. and Ketterson, John B. and Kanatzidis, Mercouri G. and Xu, Tao and Chan, Maria K. Y. and Schaller, Richard D.},
title = {Polar Fluctuations in Metal Halide Perovskites Uncovered by Acoustic Phonon Anomalies},
journal = {ACS Energy Letters},
volume = {2},
number = {10},
pages = {2463-2469},
year = {2017},
doi = {10.1021/acsenergylett.7b00790},
URL = {https://doi.org/10.1021/acsenergylett.7b00790},
}

@article{Mante2017,
author = {Mante, Pierre-Adrien and Stoumpos, Constantinos C. and Kanatzidis, Mercouri G. and Yartsev, Arkady},
title = {Electron–acoustic phonon coupling in single crystal {CH$_3$NH3PbI$_3$} perovskites revealed by coherent acoustic phonons},
journal = {Nature Communications},
volume = {8},
pages = {14398},
year = {2017},
doi = {10.1038/ncomms14398},
URL = {https://doi.org/10.1038/ncomms14398},
}

@article{Lejman2014,
author = {Lejman, Mariusz and Vaudel, Gwenaelle and Infante, Ingrid C. and Gemeiner, Pascale and Gusev, Vitalyi E. and Dkhil, Brahim
and Ruello, Pascal},
title = {Giant ultrafast photo-induced shear strain in ferroelectric {BiFeO$_3$}},
journal = {Nature Communications},
volume = {5},
pages = {4301},
year = {2014},
doi = {10.1038/ncomms5301},
URL = {https://doi.org/10.1038/ncomms5301},
}

\end{document}